\documentstyle[aps,prl,twocolumn,epsf]{revtex}
\begin{document}
\draft
\preprint{}
\twocolumn[\hsize\textwidth\columnwidth\hsize\csname @twocolumnfalse\endcsname

\title{ Determination of the phase of an electromagnetic field\\
via  incoherent detection of fluorescence}
\author{M. S. Shahriar$^{1,2}$,  Prabhakar Pradhan$^{1,2}$, Jacob Morzinski$^2$ }
\vspace{.4cm}
\address{$^{1}$Department  of Electrical and Computer Engineering, Northwestern 
Univeristy,
    Evanston, IL 60208  \\
   $^{2}$Research Laboratory of Electronics, Massachusetts Institute of Technology,
 Cambridge, MA 02139 }
\maketitle
\date{\today}
\vspace{.3cm}
\begin{abstract}
We show that the phase of a field can be determined by incoherent detection of the population of one state of 
a two-level system if the Rabi frequency is comparable to the Bohr frequency so that the rotating wave 
approximation is inappropriate. This implies that a process employing the measurement of population is not a 
square-law detector in this limit. We discuss how the sensitivity of the degree of excitation to the phase of 
the field may pose severe constraints on precise rotations of quantum bits involving low-frequency 
transitions. We present a scheme for observing this effect in an atomic beam, despite the spread 
in the interaction time. 
\end{abstract} 
\pacs{   03.67.-a, 03.67.Hk, 03.67.Lx, 32.80.Qk } 
\vskip.3cm
]
It is  well known that the amplitude of an atomic state is necessarily complex.  Whenever a measurement is 
made,   the square of the absolute  value of the amplitude is the quantity we generally measure.  The 
electric or magnetic field generated by an oscillator, on the other hand,  is real, composed of the sum of 
two  complex components.  In describing semiclassically the atom-field interaction involving such a field, 
one often  side-steps this difference by making the so-called the rotating wave approximation (RWA), under 
which only one of the two complex components is kept, and the counter-rotating part is ignored.  Under this 
approximation, an  atom interacting with a field enables one to measure only the intensity, and not the phase 
of the driving field.  This is the reason why most detectors are so-called square-law detectors.


In this article,  we show  how a single atom by itself can detect the
{\em  absolute phase}  of  a driving field, by  making use of  the 
interference between the co- and counter-rotating parts of the excitation, 
while the Rabi frequency is not negligible compared to the transition frequency.  This detection 
is performed by measuring incoherently the population of either of the two states of a two level atom.  This 
implies that a process employing the measurement of population is not a square-law detector in this limit. We 
discuss how the sensitivity of the degree of excitation to the phase of the field may enable phase 
teleportation using a pair of entangled atoms, but poses severe constraints on precise rotations of quantum 
bits involving low-frequency transitions. We also present a scheme for observing this effect in an atomic 
beam, despite the spread in the interaction time.

   We consider an ideal two-level system  where  a ground state $|0\rangle$ 
is coupled to a higher energy state $|1\rangle.$  We also assume that  the 
$0 \leftrightarrow 1$ transitions are magnetic dipolar, 
with a transition frequency $\omega$, and the  magnetic field
is  of the form $B=B_0 \cos(\omega t+\phi)$. 
We now summarize briefly  two-level dynamics without the RWA.
In the dipole approximation, the Hamiltonian can be written as:
\begin{equation} 
\hat{H} = \epsilon ( \sigma_0 -\sigma_z )/2 + g(t) \sigma_x
\label{hmt_original}
\end{equation}
where $g(t) = -g_0\left[\exp (i\omega t + i\phi)+c.c. \right] /2$, 
$\sigma_i$ are Pauli matrices, and  $\epsilon=\omega$ 
corresponds to resonant excitation. The state vector is written as:
\begin{equation} 
|\xi(t)\rangle = \left( \begin{array}
{c} C_0(t)  \\  C_1(t)
\end{array} \right). 
\label{ket_c_original}
\end{equation}
We  perform a rotating wave transformation by operating on  $|\xi(t)\rangle$
with the unitary operator $\hat{Q}$, where:                                                      
\begin{equation}
\hat{Q} = (\sigma_0 + \sigma_z )/2 +  \exp (i\omega t + i\phi)(\sigma_0 - \sigma_z)/2. 
\label{Q_operator}
\end{equation}
The Schr\"{o}dinger equation then  takes the form (setting $\hbar=1$):
$\dot{ |\tilde{\xi}\rangle } = -i H(t) |\tilde{\xi}(t) \rangle $ 
where the effective Hamiltonian is given by:
\begin{equation} 
\tilde{H} =   \alpha(t)\sigma_{+} + \alpha^{*}(t) \sigma_{-},
\label{hmt_tilde}
\end{equation}
with $\alpha(t) =  - (g_0/2)\left[\exp (-i2\omega t- i2\phi)+1 \right] $, 
and in the rotating frame the state vector is:
\begin{equation} 
|\tilde{\xi}(t) \rangle  \equiv \hat{Q}|\tilde{\xi}(t)\rangle= \left( 
\begin{array}{c} \tilde{C}_0(t)  \\  \tilde{C}_1(t)
\end{array} \right). 
\label{ket_c_tilde}
\end{equation}
Now, one may choose to make the rotating wave approximation (RWA),
corresponding to dropping the fast oscillating term in $\alpha(t)$.  This 
corresponds to ignoring effects (such as the Bloch-Siegert shift ) of the 
order of  $(g_0/\omega)$, which can easily be observable in experiment if 
$g_0$ is large \cite{Corney77,Allen75,Bloch40,Shirley65,Stenholm73,Ramsey56}.
On the other hand,  by  choosing $g_0$ to  be small enough, one can make 
the RWA for any value of $\omega$.  We explore  both regimes in  this paper. 
As such,  we find the general results without the RWA.

From Eqs.\ref{hmt_tilde} and \ref{ket_c_tilde}, one gets two  coupled differential  equations:
\begin{mathletters}
\begin{eqnarray}
\dot{\tilde{C}}_0(t) & = & - (g_0/2)\left[ 1+ \exp  (-i2\omega t-i2\phi)\right 
]\tilde{C}_1(t)\\
\dot{\tilde{C}}_1(t) & = & - (g_0/2)\left[ 1+ \exp  (+i2\omega t+i2\phi)\right] 
\tilde{C}_0(t).
\label{c_diff_eqn_b}
\end{eqnarray}
\label{c_diff_eqn_total}
\end{mathletters}
We assume  $|C_0(t)|^2=1$  is the initial condition, and proceed further 
to find  an approximate analytical solution of Eq.\ref{c_diff_eqn_total}.  Given the 
periodic nature of the effective Hamiltonian, the general solution to Eq.\ref{c_diff_eqn_total} 
can be written in the form:                                                
\begin{equation} 
|\tilde{\xi}(t)\rangle = \sum_{n=-\infty}^{\infty}
\left( \begin{array}{c} a_n  \\  b_n \end{array} \right) \exp(  n(-i2\omega t-i2\phi)). 
\label{xi_Bloch_expn}
\end{equation}
Inserting Eq.\ref{xi_Bloch_expn}  in  Eq.\ref{c_diff_eqn_total}, and equating coefficients with same 
frequencies, one gets for all $n$ :
\begin{mathletters}
\begin{eqnarray}
  \dot{a}_n & = & i2n\omega a_n + i g_0(b_n +b_{n-1})/2, \\
  \dot{b}_n & = & i2n\omega b_n + i g_0(a_n +a_{n+1})/2.
\end{eqnarray}
\label{a_n_b_n_eqn}
\end{mathletters}                 
Here, the coupling between $a_0$ and $b_0$ is the conventional 
one present when the RWA is made. The couplings to the nearest 
neighbors, $a_{\pm 1}$ and $b_{\pm 1}$, are detuned by an amount 
$2\omega$,  and so on.  To the lowest order in $(g_0/\omega)$, we can ignore 
terms with $|n|>1$, thus yielding a truncated set of six equations:                
\begin{mathletters}
\begin{eqnarray}
\label{six_eqn_a}
\dot{a}_0 & = & i g_0(b_0 +b_{-1})/2, \\
\label{six_eqn_b}
\dot{b}_0 & = & i g_0(a_0 +a_{1})/2, \\
\label{six_eqn_c}
\dot{a}_1 & = & i2 \omega a_1 + i g_0(b_1 +b_{0})/2, \\
\label{six_eqn_d}
\dot{b}_1 & = & i2\omega b_1 + i g_0 a_1/2, \\
\label{six_eqn_e}
\dot{a}_{-1} & = & -i2\omega a_{-1} + i g_0  b_{-1}/2, \\
\label{six_eqn_f}
\dot{b}_{-1} & = & -i2\omega b_{-1} + i g_0 (a_{-1} +a_{0} )/2.
\end{eqnarray}
\label{six_eqn_total}
\end{mathletters}
We consider $g_0$ to have a time-dependence of the form 
$g_0(t)=g_{0M}\left[1-\exp(-t/\tau_{sw})\right]$, 
where the switching time constant  $\tau_{sw}$
is large compared to other characteristic time scales such as $1/\omega$  and $1/g_{0M}$.  
Under this condition, one can solve these equations by employing the method of adiabatic 
elimination, which is   valid to first order in $\eta\equiv(g_0/4\omega )$. Note that  
$\eta$ is also a function of time, and can be expressed as 
$\eta(t)=\eta_0\left[1-\exp(-t/\tau_{sw})\right]$,
where $\eta_0\equiv(g_0M/4\omega )$. To solve the set of equations above, we consider  
first Eqs.\ref{six_eqn_e} and \ref{six_eqn_f}. In order to simplify these two equations 
further, one needs to diagonalize the interaction between $a_{-1}$  and $b_{-1}$.  
Define $\mu_{-}\equiv (a_{-1}-b_{-1})$ and $\mu_{+} \equiv (a_{-1}+b_{-1} )$, 
which now can be used to re-express these two equations in a symmetric form as:
\begin{mathletters}
\begin{eqnarray}
  \dot{\mu}_{-} & = & -i(2\omega+ g_0/2 ) \mu_{-} - i g_0 a_{0}/2, \\
  \dot{\mu}_{+} & = & -i(2\omega -g_0/2) \mu_{+}  + i g_0 a_{0}/2.
\end{eqnarray}
\label{mu_eqn}
\end{mathletters}
Adiabatic following then yields (again, to lowest order in $\eta$ ):
$\mu_{-} \approx -\eta a_{0}$ and  $ \mu_{+} \approx \eta a_{0}$, 
which in turn yields  $a_{-1} \approx 0 $ and  $b_{-1} \approx \eta a_{0}$.   
In the same manner, we can solve equations \ref{six_eqn_c} and \ref{six_eqn_d}, 
yielding: $a_{1} \approx -\eta b_0$ and $b_{1} \approx 0$.   

Note that the amplitudes of $a_{-1}$ and $b_{1}$ are vanishing (each proportional to
$\eta^2$ ) to lowest order in $\eta$,  and  thereby justifying our truncation of 
the infinite set of relations in  Eq.\ref{six_eqn_total}.  
It is easy to show now:  
\begin{mathletters}
\begin{eqnarray} 
\dot{a}_0 & = & ig_0 b_0/2  + i\Delta (t) a_{0} /2,\\ 
\dot{b}_0 & = & ig_0 a_0/2  - i\Delta (t) b_{0}/2, 
\end{eqnarray}
\label{bs_eqn}
\end{mathletters}
where $\Delta (t) = g^2_{0}(t)/4\omega$  is essentially the Bloch-Siegert shift.  
Eq.\ref{bs_eqn} can be thought of as  a two-level system excited by a field 
detuned by $\Delta$.  
For simplicity, we assume that  this detuning is dynamically compensated for by adjusting the 
driving frequency $\omega$.  This assumption does not affect the essence of the results to 
follow, since the resulting correction to $\eta$ is negligible. 
With the  initial condition of all the 
population in $|0\rangle$ at $t=0$, the only non-vanishing  
(to lowest order in $\eta $ ) terms in the solution of Eq.\ref{six_eqn_total}  are:\\
$a_0(t) \approx \cos( g_0'(t)t/2)$, \quad\quad $b_0(t)\approx i \sin( g_0'(t)t/2)$,\\ 
$a_1(t)\approx -i \eta \sin(g_0'(t)t/2)$,  and  $b_{-1}(t) \approx \eta \cos(g_0'(t)t/2 )$,\\ 
where \\
$g_0'(t)=1/t \int_0^t g_{0}(t)dt = g_0 \left[1-(t/\tau_{sw})^{-1}\exp(-t/\tau_{sw})\right].$

 We have verified this solution via numerical integration of Eq.\ref{c_diff_eqn_total} as shown
later.  Inserting this solution in Eq.\ref{c_diff_eqn_total}, and reversing the rotating wave 
transformation, we get the following expressions for the components of Eq.\ref{ket_c_original}:
\begin{mathletters}   
\begin{eqnarray}
C_0(t) &=& \cos(g_0'(t)t/2) - 2\eta\Sigma \sin(g_0'(t)t/2), \\
C_1(t) &=& i e^{-i(\omega t + \phi)} [ \sin(g_0'(t)t/2) +  \nonumber \\ 
       &+&2\eta\Sigma^* \cos(g_0'(t)t/2) ],
\end{eqnarray}
\label{c_total_eqn}
\end{mathletters}
where we have defined  
$\Sigma\equiv(i/2)\exp(-i(2\omega t+ 2\phi))$.  To lowest order in 
$\eta$, this solution is normalized at all times. Note that if one wants to carry 
this excitation on an ensemble of atoms using $ \pi/2$  pulse and measure the population of the 
state  $|1\rangle$  after the excitation terminates (at $t=\tau $ when 
$g'(\tau)\tau/2= \pi /2$ ), the 
result would be a output signal given by,
\begin{equation}
|C_{1}(g_0'(\tau),\phi)|^2 =\frac{1}{2}\left[1+2\eta 
\sin(2\omega\tau + 2\phi)\right]
\label{population_c1}
\end{equation}
which contains information of both the  amplitude and the phase of the 
driving  field. {\em{ This is our main result }}.

\begin{figure}
\epsfxsize=8cm
\centerline{\epsfbox{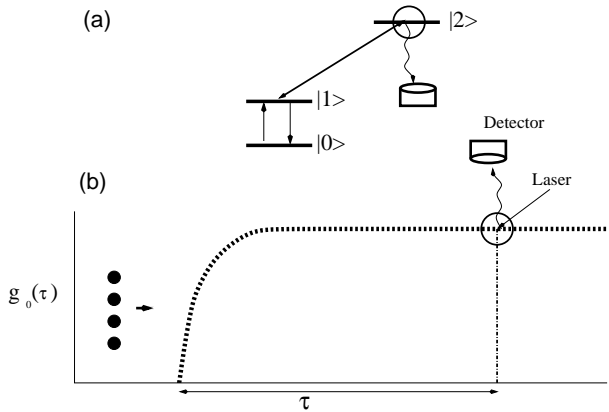}}
\caption{ Schematic illustration of an experimental arrangement for measuring the phase dependence 
of the population of the excited state $|1 \rangle$: (a) The microwave field couples the ground state 
($|0\rangle$) to the excited state ($|1\rangle$).  A third level, $|2\rangle$, which can be coupled 
to $|1\rangle$ optically, is  used to measure the population of $|1\rangle$ via fluorescence 
detection. (b) The microwave field is  turned on adiabatically with a switching 
time-constant $\tau_{sw}$, and the fluorescence is monitored after a total 
interaction time of $\tau$.}
\label{fig_1}
\end{figure}

A physical realization of this result can be appreciated best by
considering an experimental arrangement of the type illustrated in Fig.\ref{fig_1}.
Here, a single group of atoms (e.g., atoms held in a dipole force trap) are 
subjected to a resonant, oscillating magnetic field, which is
turned on  adiabatically with a switching time-constant  
$\tau_{sw}$, starting at t=0. After an interaction time of $\tau$ , 
the population of the excited state ($|1\rangle$) is determined 
{\em{instantaneously}} (i.e., with a time-constant 
much faster than $1/\omega$  and $1/g_{0M}$ ) by coupling this state to an 
optically excited state ($|2\rangle$) with a laser, and monitoring the resulting
fluorescence.  Such a measurement would correspond to 
the expression in Eq.\ref{population_c1}. 
Defining the phase of the field 
at $t=\tau$ to be  $\phi_{\tau}\equiv \omega \tau + \phi$, 
the expression of Eq.\ref{population_c1} can be rewritten as:
$ |C_{1}(\tau)|^2 =\frac{1}{2}\left[1+2\eta \sin(2\phi_{\tau})\right]. $

\begin{figure}
\epsfxsize=8cm
\centerline{\epsfbox{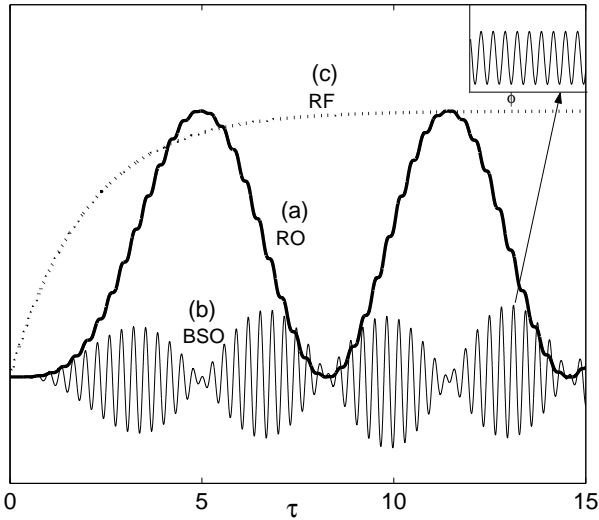}}
\caption{ Illustration of the Bloch-Siegert Oscillation (BSO):  (a)  The population of state 
$|1\rangle$, as a function of the interaction time $\tau$ , showing the BSO superimposed on the 
conventional Rabi oscillation. (b) The BSO oscillation (amplified scale) by itself, 
produced by subtracting  the Rabi oscillation from the plot in (a). 
(c) The time-dependence of the Rabi frequency. Inset: BSO as a function of the absolute 
phase of the field.}
\label{fig_2}
\end{figure}

In  Fig.\ref{fig_2}(a)  we have shown the evolution of the excited state
population  $|C_1(\tau)|^2$ as a function of interaction  time $\tau$,   
using  the  analytical expression of  Eq.\ref{c_total_eqn}(b).
Under the RWA, this curve would represent the conventional Rabi oscillation. 
However, we notice here some additional oscillations, which is magnified 
and shown separately in figure 2(b), produced by subtracting the 
conventional Rabi oscillation ($\sin^2(g(t)/2)$) from figure 2(a).
That is, figure 2(b) corresponds to what we call the Bloch-Siegert Oscillation (BSO), 
given by  $\eta \sin(g'_0(\tau)\tau)\sin(2\phi_{\tau})$.
The dashed curve (c) shows the time-dependence of the Rabi frequency. 
These analytical results agree very closely  to the 
results which are obtained via direct numerical integration of  Eq.\ref{c_diff_eqn_total}.
Note that the BSO is at twice the frequency of the driving field, and its amplitude is 
enveloped by a function that vanishes when all the atoms are in a single state.

  Consider next a situation where the interaction time,$\tau$, is fixed so that we are at 
the peak of the BSO envelope (which corresponds to a ($\pi /2$) pulse for the Rabi 
oscillation). We further assume that $\tau $  is long enough so that 
$g_0(\tau)\approx g_{0M}$.  
The experiment is now repeated  many times, with a different value of $\phi$  each time.  
The corresponding population of $|1\rangle$ is given by 
$ \eta_0 \sin(2\phi_{\tau})= \eta_0 \sin(2(\omega\tau +\phi ))$, and is plotted 
as a function of $\phi$  in the inset of figure \ref{fig_2}.  This dependence of the 
population of $|1\rangle$ on the initial phase $\phi$ (and, therefore, on the final 
phase $\phi_{\tau}$  ) makes it possible to measure these quantities.  

 As indicated above, these effects can be observed in an experiment where, for example, a 
stationary collection of atoms are excited repeatedly by a microwave field.  However, a 
more robust process for observing this effect can be realized using an atomic beam.  For 
illustration, consider first a situation where the atoms are emitted in regular intervals  
$\Delta t $, and propagate in the $z$ direction with a fixed velocity $v$.  We can describe such a 
source by the line density of the number of atoms at a position $z$ and at an instant $t$:  
$ M(z,t)=m\sum_{l=0}^{\infty}\delta \left[ z-v(t-l\Delta t)\right].$

 The microwave field is assumed to be spatially varying, corresponding to a Rabi frequency 
that vanishes for $z < z_0$.  For $z \ge z_0$, it is given by:
$ g_0(z) = g_{0M} \left[1- \exp(-(z-z_0)/z_{sw})\right]$
where $z_{sw}=v t_{sw}$ represents the distance over which the field is switched on.  
Under these assumptions, and in the limit where $\Delta t\rightarrow  0 $ 
(corresponding to a continuous, monovelocity atomic beam), it is easy to show that the 
normalized population $S$ of $|1\rangle$ measured at a 
position $z=z_o+v\tau$ (where $\tau$  is the interaction duration corresponding to 
a $\pi/2$ pulse) as a function of time is given simply by  
$[1/2+\eta_0 \sin(2\omega t+ 2 \phi)]$, assuming that the microwave field at 
this position is $B=B_0 \cos(\omega t+\phi )$.  Note that ($\omega t+\phi $) represents the 
absolute phase of the field as seen by the atoms when they arrive at this position. 
Thus, measurement of $S$ directly reveals the absolute phase of the field, modulo $2\pi$.  
Alternatively, one can mix $S$  with another signal $F$ corresponding to the second harmonic 
of a phase-shifted version of the microwave field $(F=F_0 \cos[2(\omega 
t+\phi-\pi/2- \theta)])$, where $\theta$ is a controlled, variable phase shift), 
and observe the dc component of this signal, which will be proportional 
to $\cos(\theta )$.

 Consider next the more easily realizable situation where the atomic beam is produced by an 
effusive oven.  Such a beam is typically characterized by a normalized velocity  distribution  
$f(v)=2v^3u^{-4}\exp(-v^2/u^2)$, where the $u= \sqrt{2kT/m}$ is the most probable valocity, $K$ 
is the Boltzman constant, $T$ is the temperature and $m$ is the mass of the atom \cite{Ramsey56}.
The atoms that contribute to $S$ come from all different velocity groups.  As such, each 
group experiences a different initial phase (i.e., phase at $z=z_0$), and one might think that 
this would cause the signal to wash out. However, notice that $S$ corresponds not to the initial 
phase, but rather to the phase at the observation point.  Since the atoms contributing to $S$ all 
have, by definition, arrived at this point at the same time, the signal will have the same time 
dependence, independent of the velocity group.  Explicitly, the expression for $S$ now becomes:
\begin{eqnarray}
S(t) = \int_{v=0}^\infty && dv f(v)[|\sin(g'_0(\tau_v)\tau_v/2)|^2 +\nonumber \\
          &&\eta \sin(g'_0(\tau_v) \tau_v) \sin(2\omega(t+\phi))].
\label{signal_mb}
\end{eqnarray}
where  $\tau_v =\tau u/v$ is the effective interaction time for the atoms with velocity $v$,
and  is a constant, corresponding to the interaction time for the most probable velocity $u$. 
The terms inside the integral in Eq.\ref{signal_mb} are independent of $t$ and $\phi$  , and the 
effect of this averaging is simply to reduce the amplitude of the oscillatory signal observed.

  When a quantum bit (qubit) is represented by two non-degenerate states of a massive particle, it 
is necessary to apply a field at a frequency matching the energy difference between these 
states, in order to produce an arbitrary rotation of the qubit.  In order to minimize the 
decoherence rate of such a qubit, one often chooses to use low energy spin transitions.  In 
general, one is interested in performing these transitions as fast 
as possible \cite{Bouwmeester00}.  As such, it is desirable to use
a strong Rabi frequency.  The ratio of the Rabi frequency to the qubit 
transition frequency is therefore not necessarily very small.  One example of such a situation 
is already seen to occur in qubits based on trapped ions, for example 
\cite{Steane97a,Steane97b,Jonathan92}. Under this condition, 
one can not ignore the effect of the counter-rotating term.  However, as we have 
shown here, the degree of excitation (e.g., the amplitude of the excited state) depends not only 
on the product of the Rabi-frequency and the duration of the excitation, but also on the phase 
of the field at the time the interaction stops. Thus, one has to keep track of the phase of the 
excitation field at the location of the qubit  \cite{Pradhan02a,Pradhan02b}.
In principle, the phase-tracking approach embodied in this paper itself can 
be used for this purpose. Alternatively, one has to limit the 
strength of the Rabi frequency to a level dictated by the precision required of the particular 
qubit operation involved. We note that this effect is present for both direct radio-frequency 
excitation, as well as for indirect Raman excitation, which are functionally equivalent
\cite{Thomas82,Shahriar90,Shahriar97}.
Finally, we point out that by making use of distant entanglement, this mechanism may enable 
teleportation of the phase of a field that is encoded in the atomic state amplitude, for 
potential applications to remote frequency locking 
\cite{Jozsa00,Lloyd01,Levy87,Shahriar01}.
 
In conclusion, we have shown that when a two-level atomic system is driven by a strong periodic 
field, the Rabi oscillation is accompanied by another oscillation at twice the transition 
frequency, and this oscillation carries the information about the absolute phase of the driving 
field. One can detect this phase by simply measuring only the population of the excited state. 
We have also shown how this effect may be observed using an atomic beam even if it has a 
substantial velocity spread.  Finally, we have shown how this effect has to be taken into 
account in qubit operations.
 
 We wish to acknowledge support from DARPA grant No. F30602-01-2-0546 under the QUIST
program, ARO grant No. DAAD19-001-0177 under the MURI program, 
and NRO grant No. NRO-000-00-C-0158.
~

\end{document}